# AV-PsySafe: A risk model and analysis method for the psychological safety of human and autonomous vehicles interaction

Yandika Sirgabsou[a]*, Benjamin Hardin[b], François Leblanc[a], Efi Raili[c], Pericle Salvini[b], David Jackson[c], Marina Jirotka[b], and Lars Kunze[d]

[a]*Capgemini Engineering, Toulouse, France;*

[b]*Dept. of Computer Science, University of Oxford, UK;*

[c]*Capgemini Engineering, UK;*

[d]*Bristol Robotics Laboratory, University of the West of England, UK;*

*Yandika Sirgabsou @ yandika.sirgabsou@capgemini.com; Capgemini Engineering, 4 Avenue Didier Daurat, 31700 Blagnac, France

# AV-PsySafe: A risk model and analysis method for the psychological safety of human and autonomous vehicles interaction

The rapid advancement of artificial intelligence and autonomous driving technologies has significantly propelled the development of autonomous vehicles (AVs). However, psychological barriers continue to impede widespread AV adoption, despite technological progress. This paper addresses the critical yet often overlooked aspect of psychological safety in AV design and operation. While traditional safety standards focus primarily on physical safety, this paper emphasizes the psychological implications that arise from human interactions with autonomous vehicles, highlighting the importance of trust and perceived risk as significant factors influencing user acceptance. The paper makes a methodological proposal, a framework for addressing AVs psychological safety consisting of three key contributions. First, it introduces a definition of psychological safety in AVs context. Secondly, it proposes a risk model for identifying and assessing AVs psychological hazards and risks. PsySIL (Psychological Safety Integrity Level), a classification of AV psychological risk levels is developed. Thirdly, an adapted system-theoretic analysis method for AVs psychological safety is proposed. The paper illustrates the application of the framework for assessing potential psychological hazards using a scenario involving a family's experience with an autonomous vehicle, pioneering a systems approach towards evaluating situations that could lead to psychological harm. By establishing a framework that incorporates psychological safety alongside physical safety, the paper contributes to the broader discourse on the safe deployment of autonomous vehicle, aiming to guide future developments in user-centred design and regulatory practices, while acknowledging the limitations brought by the application of the proposals on a rather simple but pedagogical illustrative example

## 1 Introduction

Safety-related systems[1] such as aircraft, trains, and cars are becoming increasingly complex, intelligent, and autonomous thanks to the application of new technologies such as Artificial Intelligence (AI). However, the current public acceptability of these systems is still uncertain as it is not guaranteed to be universally welcomed (Liu et al., 2019; Xu et al., 2018). One question that arises is how well users feel supported by these systems and how to design systems in a way that not only ensures physical safety but also promotes trust and safety from a psychological perspective. In particular, this paper focuses on this question in the context of autonomous vehicles (AVs), where the largest roadblocks to AV mass adoption may now be psychological, not technological (Shariff et al., 2017; Xu et al., 2018). For instance, trust and perceived risk have been widely shown to be two major concerns of AVs (Brell et al., 2019; Kenesei et al., 2022; Li et al., 2019; Thomas et al., 2020). Individuals often fear autonomous vehicles, but they may choose to use the technology despite the fear because of its provided benefits (Cugurullo & Acheampong, 2024). However, this potentially introduces a situation of psychological risk where many people will use a safety-related system in their day-to-day life which they fear.

We argue that the inherent relationship that one develops with an AV is one of vulnerability. The safety of one's life is placed in control of the autonomy (the ability of a system to make decisions and react to unforeseen events without the human's intervention), and this high-risk relationship leads to high psychological stake. This relationship and its stake mean that it will be important for future AV developers to consider how psychological hazards introduced by the autonomy can be mitigated, whether singular catastrophic events or repeated negative exposure. Hence, the question must be asked of how to ensure that AVs are designed in a way that also takes into consideration the psychological risks stemming from human and autonomous vehicle interactions. To systematically evaluate psychological hazards, the foundations must be laid for the analysis.

[1] A Safety related system is a system whose failure or malfunction may result in death or serious injury to people and/or significant property damage or environmental harm. Such systems can be found in various industries, e.g. aerospace, medical, nuclear, automotive.

Ensuring safety in vehicles requires standardised engineering processes and methods (*ISO 26262 [2018] Road Vehicles-Functional Safety-Part 1*, 2018). In relation to AVs,  safety related approaches to guide systems design have been proposed for this purpose. Such approaches are described in recent automotive safety standards including the SotIF (*ISO 21448:2022 - Road Vehicles — Safety of the Intended Functionality*, n.d.) and ISO/PAS 8800:2024 (Road Vehicles – Safety and artificial intelligence) to encompass non-deterministic behaviour that is common of AI-enabled systems such as AVs. However, these standards only address safety from a physical perspective (preventing physical harms or property damage, referred thereafter as physical safety).

In order to address the important psychological components of critical systems and human interaction like trust and vulnerability previously discussed, it is necessary to first define adequate methods for the analysis, just as they are defined for physical safety. Unfortunately, in AVs development context, there is no established methodology or framework that systematically evaluates psychological risks and translates their mitigation into specific functional requirements during the software or system development process. Although human factors considerations related to Situation Awareness (SA) or Mental Workload (MWL) are addressed through existing methods including the Situation Awareness Global Assessment Technique (SAGAT) for SA and the NASA Task Load Index (NASA-TLX) for MWL, these methods retain a focus on physical safety and systems performance. Methods have typically been open-ended, encouraging designers to explore risks without a structured evaluation system, have focused solely on evaluating a user or worker's mental health after human-robot interaction, or only target one component of psychological safety such as stress (Winfield et al., 2023; Lee & Lim, 2025; Alenjareghi et al., 2024; Zacharaki et al., 2020). These are valuable methods, but do not offer concrete directions for designers to examine their system during the development process and systematically and extensively locate areas of psychological risk to address. Hence, from a systems safety perspective, and consistent with the general objective of the paper (which is ensuring psychological safety within the context of autonomous vehicles), we propose AV-PsySafe, a framework for human-AVs psychological assessment. This framework consists of:

1. A theoretical definition and scope of human-AV psychological safety,
2. A psychological safety risk model,
3. A psychological hazard analysis method

With a key aspect of the paper being the novelty of psychological risk in the systems context, the paper does not claim a full validation of the proposals. Instead, it illustrates the relevance and applicability of our theoretical foundations through a simple applicative example built around a driver's experience operating a new autonomous vehicle in a highway context. This scenario allows us to use our proposed psychological risk model to identify potential psychological hazards and demonstrate the use of our methodology for highlighting system design issues that lead to these hazards Likewise, the contribution of this paper does not intend to estimate how human psychological risk perception can affect driving characteristics and performance (i.e. takeover behaviours, following distance, etc.) which have been previously studied (Bellem et al., 2018; Ma & Zhang, 2021; Wang et al., 2020). Although ensuring psychological safety will undoubtedly have a positive effect on driving characteristics and performance, the focus of the paper is on preventing psychological risk in other to promote trust. The remainder of the paper is structured as follows. The next section (2 ) outlines the state of the art in psychological safety, systems safety, their principles their concepts. Section 3 details the contributions of the paper, illustrated through the example scenario. Section 4 discussed the results and their limitations. Section 5 concludes the paper.

# 2 State of the art

## *2.1 Psychological safety*

As we consider how to add the human element to risk evaluation, we can look to the well-established notion of psychological safety. Psychological safety has been defined in the occupational (workplace-related) health context as "a shared belief held by members of a team that the team is safe for interpersonal risk taking" (A. Edmondson, 1999). Thus, psychological safety considers not only direct psychological harms caused by another party, but also beliefs about the other party's trustworthiness to do what they say and perform to a high standard. Psychological safety "involves but goes beyond trust", being a distinct concept from trust which affects various behavioural and organizational results (A. Edmondson, 1999; A. C. Edmondson, 2004).

The notion of psychological safety has also been discussed in the context of physical human-robot interaction (pHRI) (Bauer et al., 2008; Lorenzini et al., 2023; Akalin et al., 2022; Kamide et al, 2012). In this context, psychological safety around robots can be understood as "the expectation that humans perceive interaction with robots as being safe, and that the interaction does not result in any psychological discomfort or stress" (Lasota et al., 2017). This motivates the question of how psychological discomfort can be defined and identified, particularly compared to the expected psychological discomfort from interaction with a non-automated version of a system (Kamide et al., 2013), and how the risk of such unacceptable harm can be addressed.

Psychological safety of autonomous systems is closely related to the notions of perceived safety which can be a determining factor for trust Click or tap here to enter text.(Akalin et al., 2022; Kamide et al, 2012, Naiseh et al., 2024) and is influenced by a user's mental model of system functionality, testing, and reliability (Othman et al., 2023). Therefore, psychological safety derives from perception and feelings resulting from physical interactions, in contrast to the objective and deterministic nature of physical safety (where risk of harm is quantified as a product of severity and likelihood). For instance, vehicle behaviour that is completely safe but unusual, and whose intention is not well communicated to humans, can be perceived as dangerous or even a threat. Furthermore, risk of psychologically negative interactions may increase if the user perceives the system as another being or as having cognitive abilities (Winkle et al., 2021).

### *2.1.1 Psychological safety in AVs*

Prior work has shown that these psychological considerations are consequential, with trust being a significant predictor of intention to use AVs, perceived usefulness being a strong predictor of willingness to re-ride, well-being affecting intention to use an AV, and perceived risk, social trust, and perceived benefit being predictors of willingness to pay for an AV (Liu et al., 2019; Meyer-Waarden & Cloarec, 2022; Xu et al., 2018). Despite these concerns, psychological safety has not been formally defined for AVs, leading us to propose a theoretical definition in this paper. Moreover, despite psychological safety being recognised as a key factor for the adoption of AVs, it is out of the scope of current systems safety frameworks which lack methods to support the adoption component (Naiseh et al., 2024).

### 2.1.2 *Psychological risk assessment and models*

In psychology, psychiatry, and legal contexts, psychological injuries are assessed and classified thanks to well-known scales. These include self-reported questionnaires such as the Psychological Injury Risk Indicator (PIRI) (Winwood et al., 2009), the Kessler 10 (K10) and embedded Kessler 6 (K6) distress scales (Kessler et al., 2002), or the Impact of Event Scale-Revised (IES-R). However, these scales tend to only focus on the severity aspect of the psychological injuries, overlooking other components of the risk such as the probability or controllability. Moreover, these scales are used to assess psychological harms that have already occurred in individuals. However, managing risk in systems whose interaction with humans would result in psychological harm requires assessing the system under design for hazardous behaviours (i.e. potential source of ham) in different contexts in a systemic and proactive manner (as early as during the system concept and system design).

## 2.2 *Systems safety*

Safety (in a physical sense) is a concept well-mastered in systems engineering, where it has been commonly defined as “freedom from unacceptable risk”, where risk is a combination of likelihood of a harm occurring and its potential severity (usually captured in terms of injuries or death). Systems safety applies specialized knowledge, engineering and management principles, criteria, and analysis methods and techniques to achieve acceptable mishap risk within the constraints of operational effectiveness, suitability, time, and cost throughout all phases of the system life cycle (DoD, 2012). Such techniques include hazard analysis and risk assessment activities that are well defined by standards, planned, and conducted as an integral part of systems lifecycle through classical safety analysis methods such as Fault Tree Analysis (FTA) (Roberts & Haasl, n.d.) and Failure Modes and Effects Analysis (FMEA) (SAE International, 2021). Each of the safety analysis techniques that are used in traditional safety, deductive (e.g. FTA) or inductive (e.g. ETA), have strengths and weaknesses. The choice of method generally depends on what is being addressed, the type of system, and the stage of lifecycle. Usually, safety engineering will use a mixture of deductive and inductive techniques (or otherwise called top-down and bottom-up techniques) to ensure completeness of coverage of identified issues. Complementary to classical safety techniques, Human Factors (HF) related safety analyses include the Human Reliability Assessment (HRA) (NASA Chang, 2006) and the human factor Safety Critical Task Analysis (SCTA) (Energy Institute, 2020). The focus of these techniques is to assess how relevant HF aspects such as human errors or cognitive overload may contribute to system hazards. Nevertheless, the consideration of these HF aspects is still from the perspective of ensuring physical safety.

In addition to these emerging HF aspects, concerns stemming from the growing complexity and the increased reliance on embedded software have driven the development of additional safety techniques such as the System-Theoretic Accident Model and Processes (STAMP) and the System-Theoretic Process Analysis (STPA) (N. G. Leveson, 2012). STAMP is an accident causation model (i.e. theoretical foundation for hazard analysis and risk assessment). It is based on system thinking and control theory concepts (Leveson, 2004). In contrast to classical safety analysis techniques that rely on chain of events models, accident models based on system theory like STAMP consider accidents as arising from the interactions among system components (Rasmussen, 1997). In STAMP, accidents are conceived as resulting not from component failures, but from inadequate control or enforcement of safety-related constraints on the design, development, and operation of the system (N. Leveson et al., n.d.). Consequently, safety is viewed as a control problem (i.e. accidents occur when component failures, external disturbances, and/or dysfunctional interactions among system components are not adequately handled) and addressed through the adequate enforcement of system constraints, hierarchical control, and process models. STPA is an approach to hazard analysis based on STAMP (N. Leveson & Thomas, 2018). It assists safety analysts and system designers in identifying hazards and the set of scenarios that can lead to an accident and specifying safety-related constraints that must be enforced.

Beyond physical safety, STAMP and STPA allow for the consideration of other system properties deemed important by the systems stakeholders. Thus, this approach makes provisions for addressing other types of non-physical risks (such as financial, reputational, or psychological) defined as losses deriving from the violation of stakeholders needs. STAMP and STPA provide particular benefits when dealing with emergent properties in complex sociotechnical systems that include human and software elements. Through the use of a hierarchical control structure, STAMP allows the system (interrelated components) to be seen as a dynamic control loop where emergent issues are captured in the context of system component failures, external disturbances, and/or dysfunctional interactions among system components (including human and software) which are not being adequately handled. This makes the approach a suitable candidate for the assessment of emergent issues such as trust or perceived safety in AVs.

# 3 AV Psychological Framework (AV-PsySafe) Proposal

## *3.1 Research Method*

Considering the high vulnerability and trust stakes of AVs, a general problem concerned with ensuring human-AVs interaction psychological safety was formulated: **"How can AVs be designed** with consideration for **psychological risks stemming from human and autonomous vehicles interactions in order to promote trust and psychological safety?"** With AVs being sociotechnical systems, a hypothesis is made to adopt a systems engineering safety approach, leading to the formulation of the general research question: **"How can the psychological safety of human-AVs interactions be ensured in a systems engineering safety approach?"** In the light of systems safety principles, an assertion was made that answering this general research question requires 1) defining psychological safety in human and AVs interaction, 2) defining its risk model, and 3) defining a method for its assessment in AVs context. Consequently, the following three (3) specific research questions were formulated.

1 **How can psychological safety be defined in human-AV interactions to enable the analysis of psychological risk?**

2 **How can existing safety theoretical models and concepts be applied to create a comprehensive framework for conceptualizing psychological risks in human-AV interactions?**

3 **How can existing safety analysis methods and techniques be applied or adapted to enable the identification, analysis, and mitigation of psychological risk?**

### *3.2 Illustrative example*

Let's consider a driver's first experience aboard an SAE Level 4 autonomous vehicle in a highway driving lane change scenario. According to (*Taxonomy and Definitions for Terms Related to Driving Automation Systems for On-Road Motor Vehicles*, 2021), a Level 4 vehicle is capable of handling all driving tasks under specific conditions without requiring human intervention, including responding to unexpected events and failures. In this scenario, let's imagine that while our autonomous vehicle approaches an interchange, another vehicle from the right engages in a manoeuvre to merge onto the same lane as the AV, as illustrated in Figure 1. In such an event, the human driver would activate the left turn indicator, slightly adjust speed, if necessary, then gently move to the left when the lane is clear. Once the third-party vehicle had entered the lane, our driver would check the traffic coming from behind, activate the right turn indicator, and pull back into the original lane. However, as the 3$^{rd}$ party vehicle enters the merge, the AV engages in an unexpected behaviour by alternatively swerving left to right while pulling to the left, as illustrated in Figure 1. Realizing that something might be wrong, the human requests takeover. His request is denied, as the Automated Driving System (ADS) manages to complete the manoeuvre while simply alerting the user of a braking system malfunction and stating that the ADS is attempting to correct the issues. However, the message is so generic that the human driver is unable to understand it in the limited amount of time that the driver must process the message. After the third-party vehicle had entered the lane, the ADS attempts to pull back to the initial lane. As it does so, the same unexpected swerving behaviour occurs again, as illustrated in Figure 1.

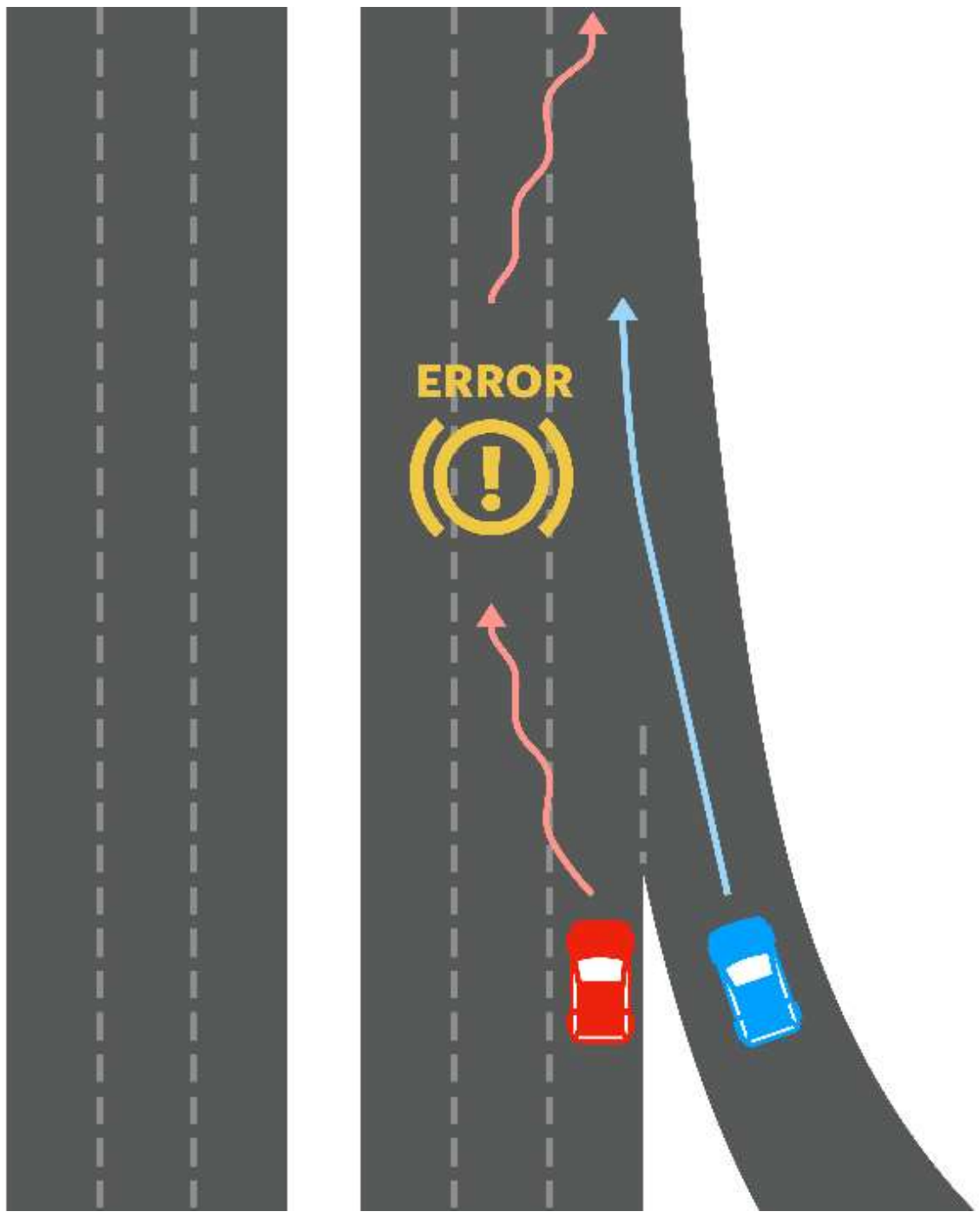

Figure 1. Illustrative example of an AV's tactical manoeuvre malfunction. Ego vehicle (red) experiences a braking malfunction during a lane change related to another vehicle (blue) merging onto the motorway.

One might intuitively recognise several factors in this scenario that puts this driver at psychological risk, but how can we be sure these factors are systematically identified and addressed during vehicle design? This illustrative example will be used throughout the paper, and we limit this paper to this single, straightforward example as a way to demonstrate the components of the analysis. Once the framework has been presented, future work can further validate its applicability across diverse driving scenarios, potentially based on real-world data. ***In the remainder of the section, we present our contribution of AV-PsySafe, describe how it has grown out of the existing safety methods, and consider results of its use applied to the illustrative example.***

### *3.3 Contribution and Results*

#### *3.3.1 Theoretical definition of AV psychological safety and the scope of our analysis*

Ensuring psychological safety in human-AV interactions context necessitates establishing a clear definition of what it is. As revealed in the state of the art (Section 2.1), psychological safety is complex, is highly dependent on perceptions and impressions, is not necessarily aligned with objective and empirical risk, and occurs in an interactive context (human interaction with the vehicle) (Lasota et al., 2017). Thus, drawing from the previously introduced definitions by both (Lasota et al., 2017) in robotic psychological safety and (A. Edmondson, 1999) in occupational psychological safety, we propose the following definition for autonomous vehicle psychological safety..

**Definition.** Psychological safety in AVs is defined as:

> *The absence of unacceptable psychological risk resulting from autonomy use or misuse, where psychological risk is a combination of the probability of occurrence of psychological harm and the severity of that harm.*

**Components of psychological safety.** Drawing on extensive literature investigating human factors in AVs, we characterize AV psychological safety as being comprised of several components including trust, perceived control, perceived risk, psychological empowerment, responsibility and liability, comfort, predictability, perceived support, effort-reward, appearance, and low demands (Brell et al., 2019; Guettas et al., 2019; Harvey et al., 2017; Hoff & Bashir, 2015; Wang et al., 2020). One can almost consider psychological safety as an umbrella term for the previously explored components, however, it aims to tie these together in a usable way, ensuring that their individual influences are recognised, as we will later demonstrate. Finally, (Harvey et al., 2017) examines the stress resulting from role ambiguity and role conflict, which may play a role in human-AV teamwork. Psychological empowerment and emotional exhaustion have also been shown to be influencing components of psychological safety (Zhou & Chen, 2021).

**Timeframe.** This analysis is most suited to immediate psychological risks; however, it is intended to be adaptable to long-term risks. Immediate risks are easier to identify and measure, such as the psychological response to a sudden braking manoeuvre. Nevertheless, it has been shown in other domains that psychological effects can also compound over many minor negative situations, leading to effects that develop much later and can be highly individual (McHugh et al., 2018; Udwin et al., 2000).

**Stakeholders.** The scope of this analysis is aimed primarily at the driver or overseer of an AV but is valid for other possible stakeholder interactions with the vehicle. Most notably, the analysis should be practicable for passengers of an AV who are not responsible for overseeing the operation of the autonomy. The analysis can also extend to other road users and vulnerable road users that will have their own mental model of the vehicle's automation and maintain perceptions of the vehicle, albeit from a different perspective.

**Level of autonomy.** The scope of this analysis is aimed at SAE Level 2 to Level 5, which are 4 of the 6 levels of driving automation ranging from no driving automation (Level 0) to full driving automation (Level 5) (*Taxonomy and Definitions for Terms Related to Driving Automation Systems for On-Road Motor Vehicles*, 2021). Each level of autonomy will be accompanied by different mental models and expectations; however, our analysis still applies across autonomy levels where the system is expected to perform some or all functions of driving.

**Baseline psychological state.** Each stakeholder will have a different psychological state before interacting with the autonomous system. One example state is situational anxiety. During one interaction, the individual may begin with a state of low situational anxiety, while during another interaction, they may begin with a state of high situational anxiety. Because humans perceive threats differently based on their psychological state, different individuals can have varying reactions to the same negative incident (Bennett et al., 2019; Kavcıoğlu et al., 2021; Muris et al., 2003; Udwin et al., 2000). For this reason, we assume that the presented risk analysis applies to individuals within an average range of psychological state (e.g. low to moderate situational anxiety, no prior trauma from AV interaction, average confidence level in the AV). It may be impossible to ensure system safety for individuals of all backgrounds, and thus we must limit the scope of stakeholders to the above conditions.

**Mitigating psychological risks vs. improving psychological state.** In theory, the psychological state of an individual could be improved such that risks are not only mitigated, but the user is given a more positive state than their baseline (Koch et al., 2021, Zepf et al., 2019; Paredes et al., 2018). Finding and designing these interventions is certainly an important area of future research but is outside the scope of this work. Here, we aim to assess risk and mitigation for risk.

### *3.3.2 Proposed psychological safety risk model based on STAMP*

All hazard analysis and risk assessment techniques rely on a risk model which requires defining the risk and its underlying mechanisms and mitigating risk in systems approach which first means designing systems in a way that prevents the risks. Therefore, it is necessary to understand components of psychological risk and how they occur in a given context. This is the objective of the proposed psychological risk model. It adapts key concepts from STAMP and classical safety in the automotive domain in order to define concepts necessary for psychological safety analysis in a systems approach as shown in Figure 2. Important to note that this model is not aimed to diagnose psychological outcomes from events but rather highlight unsafe control actions that could lead to potential psychological risk.

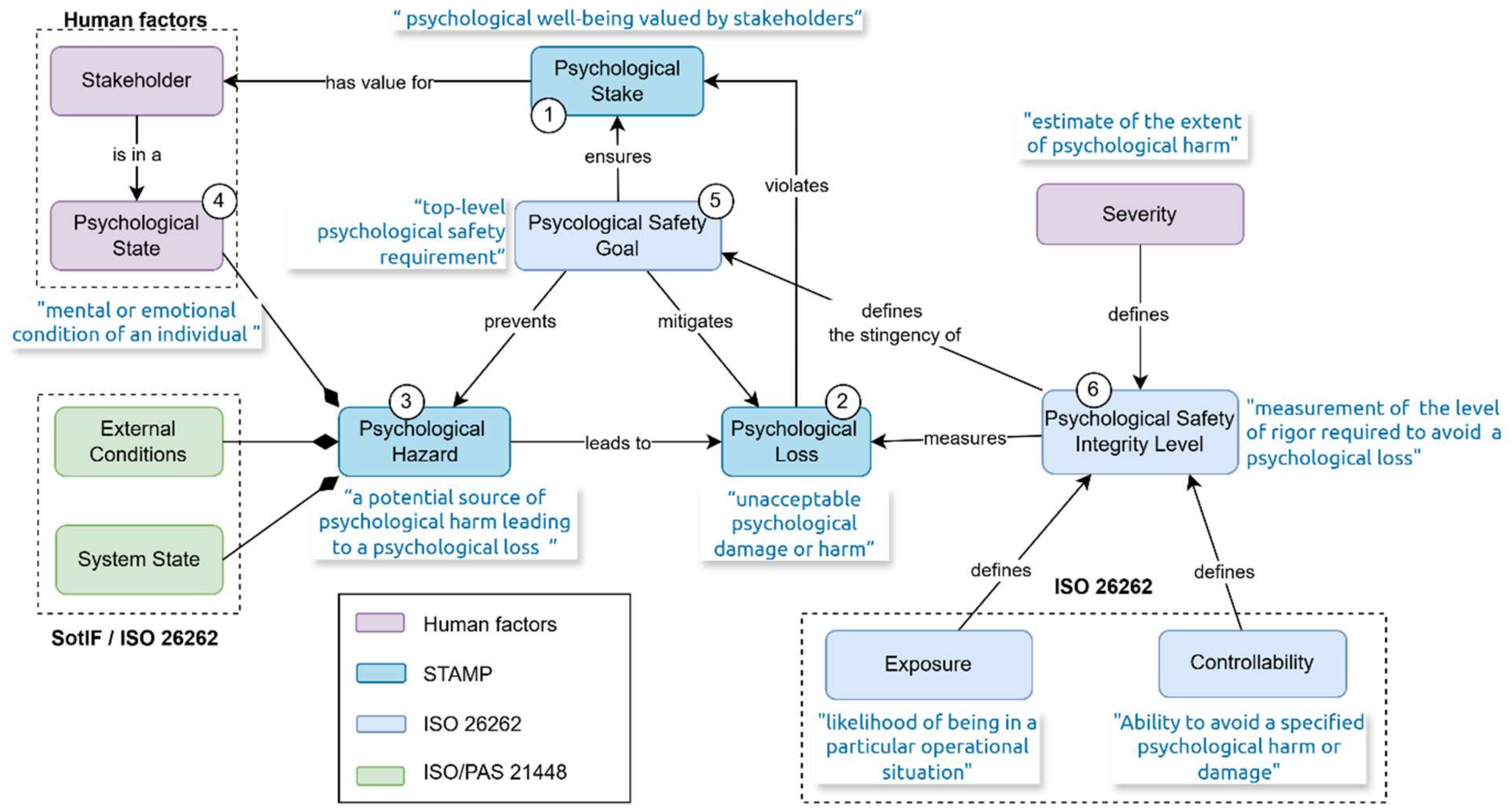

Figure 2. Proposed psychological safety concepts summarized as follows: Psychological stakes are the psychological needs or well-being of stakeholders, and their violation leads to psychological losses. Psychological hazards are potential sources of psychological losses, influenced by the psychological state of the individual, the state of the system, and external conditions. To prevent these hazards and mitigate losses, psychological safety goals are established, and the stringency of these goals is determined by our proposed Psychological Safety Integrity Level (PsySIL), considering exposure, controllability, and severity of the losses.

The first concept of our model is “psychological stake” defined as "a psychological need or well-being valued by the stakeholder(s)", such as feeling safe. As illustrated in Figure 2, the violation of these psychological stakes leads to psychological losses, defined as “any unacceptable psychological harm to stakeholders resulting from their interaction with the system” (see ① and ② in Figure 2).

In our scenario, the driver has the psychological stakes of feeling safe, feeling protected, or ensuring protection of passengers under their responsibility (if applicable), and trusting the Automated Driving System (ADS). In this case, a simple example of psychological loss is the loss of trust in the ADS. Further examples of psychological losses are detailed in Section 3.3.

The next concept is “psychological hazard”, defined as “a potential source of psychological harm/loss”. In our risk model, it is associated with a “psychological state (mental or emotional condition of an individual, workload, attitudes about the automation, and trust propensity)” (Martin, 1990; Pak & Rovira, 2024), and a particular context defined by a state of the system and a set of external conditions. See ③ and ④) in Figure 2.

With the vehicle operating under an SAE Level 4 autonomous mode, examples of psychological hazards could include the following: 1) the vehicle performs an unexpected sudden driving manoeuvre without warning the passenger when responding to an external situation (solicitation), or 2) the vehicle performs an action that causes discomfort to passenger due to unnatural motion characteristics (e.g. as a result of failure or performance limitation).

To enable the control of psychological hazards, the proposed risk model introduces the notion of “psychological safety goal” inspired from ISO 26262, which is defined as “a high-level requirement specifying the conditions and reinforcement of the system to be met in order to prevent psychological hazards and mitigate psychological losses” (see ⑤ in Figure 2). In our example, the vehicle performs an unexpected sudden driving manoeuvre when responding to an external situation (solicitation). A psychological safety goal to prevent the loss of trust resulting from this scenario would state: “The vehicle must properly inform the human driver when performing unexpected sudden driving manoeuvre when responding to an external situation.”

The risk model introduces the Psychological Safety Integrity Level (PsySIL), inspired from the Automotive Safety Integrity Levels (ASIL) (ISO 26262- Part 9:2018). PsySIL is a criterion for determining the stringency with which psychological safety goals must be met, proportional to the magnitude of the psychological losses. Similar to ASIL, PsySIL ranges from A (the least stringent) to D (the most stringent), determined by three parameters: severity, exposure and controllability as illustrated in Table 1. Unlike ASIL, where the severity is defined in terms of injury or death, PsySIL severity is defined in terms of psychological injury measured using an existing psychological risk scale based on (Taibi et al., 2022). Relying on this scale, PsySIL declines severity into three classes (marginal, moderate or critical). The exposure parameter (E) in PsySIL expresses the probability of being in an operational situation that can be hazardous if coincident with the failure mode under analysis (ISO 26262- Part 3). PsySIL determination considers 4 levels of exposure, including E1 (Very Low), E2 (Low), E3 (Medium) and E4 (High). The controllability parameter (C) expresses the ability of the autonomy to avoid causing a psychological harm or damage (e.g. by providing timely and transparent explanations), or the ability of the involved person to cope with the situation, possibly with support from the autonomy (e.g. achieving a minimal risk condition or allowing takeover by user). To determine PsySIL, the following three class of controllably are considered: C1 (simply controllable), C2 (normally controllable) and C3 (difficult to control or uncontrollable). Table 1. Determination of PsySIL, defined by exposure, controllability, and severity like in ASIL, except that the severity (marginal, moderate or critical) which is of psychological harm is derived from how long the psychological effect lasts (short, medium or long term) based on a psychological risk scale (Taibi et al., 2022).

| **Severity (S)** | | **Exposure (E)** | **Controllability (C)** | | |
|---|---|---|---|---|---|
| **Psychological Effect** | **Severity Class** | | **C1 (Simple)** | **C2 (Normal)** | **C3 (Difficult)** |
| **Short Term** (e.g. Increased heart rate, Increase in blood pressure, Adrenaline release) | **S1 (Marginal)** | E1 (Very Low) | | | |
| | | E2 (Low) | | | |
| | | E3 (Medium) | | | PsySIL A |
| | | E4 (High) | | PsySIL A | PsySIL B |
| **Medium Term** (e.g. Psychological strain, Psychosomatic health symptoms) | **S2 (Moderate)** | E1 (Very Low) | | | |
| | | E2 (Low) | | | PsySIL A |
| | | E3 (Medium) | | PsySIL A | PsySIL B |
| | | E4 (High) | PsySIL A | PsySIL B | PsySIL C |
| **Long Term** (e.g. depression, anxiety, cardiovascular disease) | **S3 (Critical)** | E1 (Very Low) | | | PsySIL A |
| | | E2 (Low) | | PsySIL A | PsySIL B |
| | | E3 (Medium) | PsySIL A | PsySIL B | PsySIL C |
| | | E4 (High) | PsySIL B | PsySIL C | PsySIL D |

Referring to Table 1, if we examine our example scenario and consider the psychological stress induced by the malfunction during the lane change, we could deduce the following:

- Exposure Level: High (E4) due to the commonality of lane changes in highway driving.
- Controllability: Simple (C1) with multiple manoeuvre options by either the user or autonomy.
- Severity: Moderate (S2), not critical, as the stress would not likely lead to severe psychological outcomes like depression or anxiety.

Based on this assessment, this scenario would result in a low-level rating PsySIL (PsySIL A).

### *3.3.3 Proposed psychological safety analysis method*

Applying the concepts described earlier in our risk model (hazards, losses, etc.), we propose to adapt STPA by focusing on psychological safety. This adaptation addresses the lack of a defined method for psychological hazard and risk analysis in systems context. This implies performing the four steps of the STPA methodology to identify psychological hazards and losses, specifying appropriate control measures to ensure the fulfilment of psychological stakes. The proposed adaptation (Figure 4) starts by defining the purpose of the analysis (Step 1), followed by the construction of the psychological safety control structure and assignment of responsibilities (Step 2). The third step identifies psychologically Unsafe Control Actions (UCA), and Step 4 identifies psychological loss scenarios. Details of each the 4 steps are described further in the following subsections along with illustrative examples.

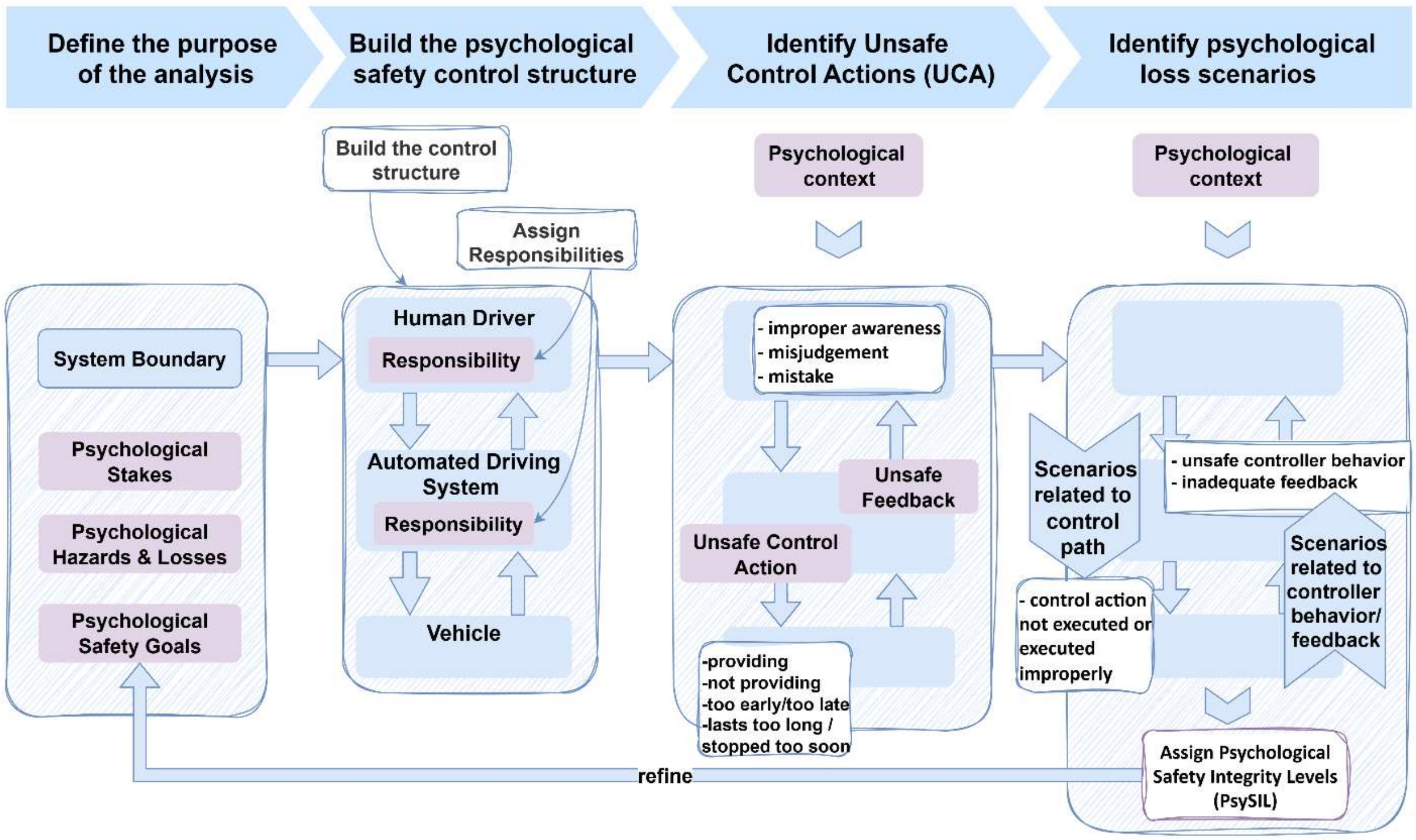

Figure 3. STPA adaptation for psychological safety analysis.

Step 1: Define the purpose of the analysis

Defining the purpose of the analysis necessitates establishing the system boundary, which distinguishes what is internal to the system under analysis and what is considered part of the environment. Defining the purpose of the analysis also includes identifying the psychological losses we want to prevent as well as identifying psychological hazards potentially leading to those losses. The step completes with the specification of psychological safety goals intended to prevent or mitigate psychological hazards and losses.

**Define the system boundary**. In our example, the system is comprised of the human driver, the ADS controller (which is the part of the Automated Driving System (ADS) responsible for executing control algorithm), and the motion control sub-systems (longitudinal and lateral), as well as interfaces (datalink and physical) between these elements. To simplify our scenario analysis, we exclude the other vehicle passengers and consider only the driver to be within the system boundary. External elements such as the road, traffic signs, and other road users (other vehicles, pedestrians, animals) are excluded from the system and considered to be part of the environment as illustrated in Figure 5. With the system boundary defined, psychological safety hazards and losses can be identified.

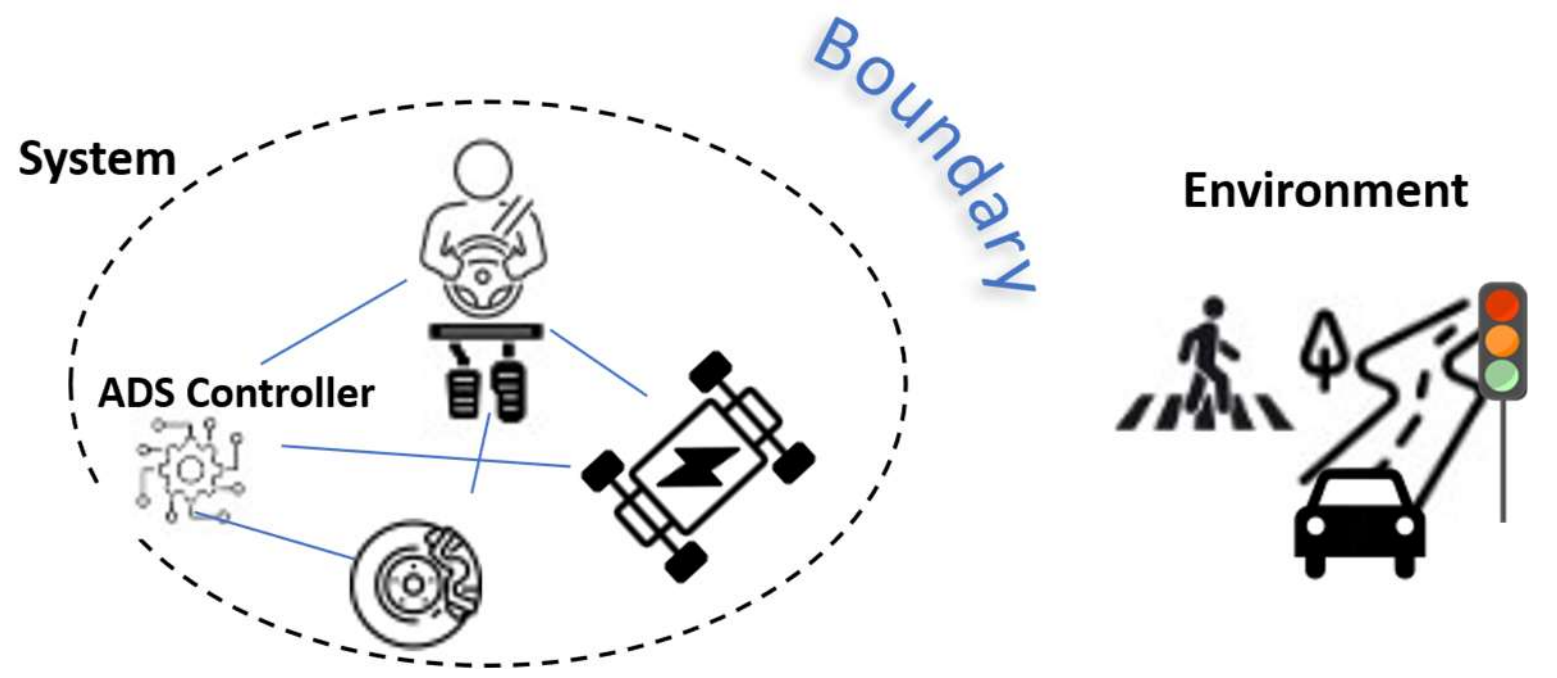

Figure 4. System boundary delimitation.

**Identify psychological losses and hazards.** Losses must be derived from the violation of stakeholder stakes. In our simple example, the identified stakeholder under consideration is the human driver. Their stakes may include trusting the automated vehicle, feeling safe (i.e. psychologically), being safe, ensuring the physical safety of the passengers. Hence, the 3 top level psychological losses deriving from the violation of these stakes are listed below:

- L1 Loss of trust.
- L2 Stress, shock, or trauma.
- L3 Fear of loss of life, injury, or property damage.

The identification of the psychological losses enables the identification of the following examples of system level psychological hazards as listed below.

- H1 - ADS Controller performs sudden tactical driving manoeuvre without informing human driver (H1 leads to L2).
- H2 - Vehicle deviates from expected behaviour when performing Dynamic Driving Task (DDT) (H2 leads to L2).
- H3 - ADS Controller ignores human driver requests (e.g. takeover, emergency stop) (H3 leads to L1, L2 and L3).
- H4 - Vehicle performs DDT while out of Operational Design Domain (ODD) (H4 leads to L3).
- H5 - Human driver misinterprets information from ADS Controller (H5 leads to L1 and L2).

As the list demonstrates, one hazard can be related to more than one loss. In our scenario, H2 occurs when the vehicle experiences an unexpected behaviour during the DDT (uneven and erratic braking). This raises the likelihood for both L1 and L2 psychological losses, as the human driver may both lose trust in the braking system (L1) and be shocked at the behaviour, elevating their stress (L2).

**Specify psychological safety goals.** Psychological safety goals express what the system must do to prevent psychological hazards or mitigate associated psychological losses, and we derive these goals from the previous hazard identification. In the context of our example, the following psychological safety goals are specified:

- SG1 - ADS Controller must properly inform the human driver when performing a sudden emergency DDT manoeuvre (SG1 prevents H1).

- SG2 - The vehicle must perform DDT manoeuvre in a manner that causes least stress to the human driver (SG2 prevents H1, H2 and H5).
- SG3 - ADS Controller must comply to human driver request (takeover, emergency stop) (SG3 prevents H3).
- SG4 - Vehicle must comply with ODD specification (SG4 prevents H4).
- SG5 - ADS user must monitor and understand the state of DDT performance (SG5 prevents H5).

As an example of one psychological loss mitigation in our scenario, the vehicle could have provided more informative warnings of vehicle behaviour, possibly a post-hoc explanation, and a description of if maintenance is needed, similar to common explainability methods in autonomous driving research (Kim et al., 2023; Omeiza et al., 2022).

Step 2: Model the psychological safety control structure

The second step focuses on constructing the psychological safety control structure that will be used for the psychological safety analysis. In STAMP, a control structure is an abstract model of the system that captures functional relationships and interactions between system components modelled as a set of feedback control loops (N. G. Leveson, 2012). We propose a control structure arranged in three levels of hierarchy: the human driver, the ADS controller, and the vehicle, which is the controlled process, as shown in Figure 5.

**Build the control structure.** For the "human driver" part, the control structure refined the mental model elements proposed in STAMP by considering components of Situation Awareness (SA) categorized in 3 levels (perception, comprehension, projection) based on (Endsley, 2015) and the human driver's beliefs and "psychological state". Referring to the "human driver" part of the control structure in Figure 5, one can understand that improper SA (e.g. misjudgement) may lead to erroneous "beliefs" about either the status of the system or external events, resulting in both a negative psychological state (e.g. as stress, fear, or loss of trust) and an inadequate control decision. Note that the proposed control structure allows for the consideration for individual factors such as age, gender, or experience since these factors affect both SA and what the users believe about the system's state. For instance, when facing an unexpected situation, an experienced AV driver may be able to better project and anticipate the ADS reaction, which would correspond to a level 3 SA, whereas an unexperienced AV drivers SA in the same situation could be limited to level 1 (perception). In the proposed psychological control structure, the human driver is responsible for making decisions and transmitting their goal, need, or intent to the ADS controller.

The vehicle and ADS controller components of the control structure were built based on the SAE functional specification related to driving automation defined in SAE J3016 (*Taxonomy and Definitions for Terms Related to Driving Automation Systems for On-Road Motor Vehicles*, 2021) that specifies the functional architecture of driving automation, listing its functional components, their dependencies, and the intended behaviour (Serban et al., 2020). Using these specifications, we developed the control structure in Figure 5. According to this model, the ADS is responsible for performing all strategic planning, tactical manoeuvring, and basic operational vehicle motion control based on the human driver's request to which it shall also provide feedback information such as warnings, system state, or certain events. The information feedback enables the human driver to form their belief about the current state of the system based on their SA level.

The vehicle sits at the lowest level of the hierarchical control structure and executes the control commands from the ADS controller or human driver. Through sensing, the ADS controller collects necessary data related to external objects and vehicle state and uses this information to form the environment models and the controlled process used by its control algorithm.

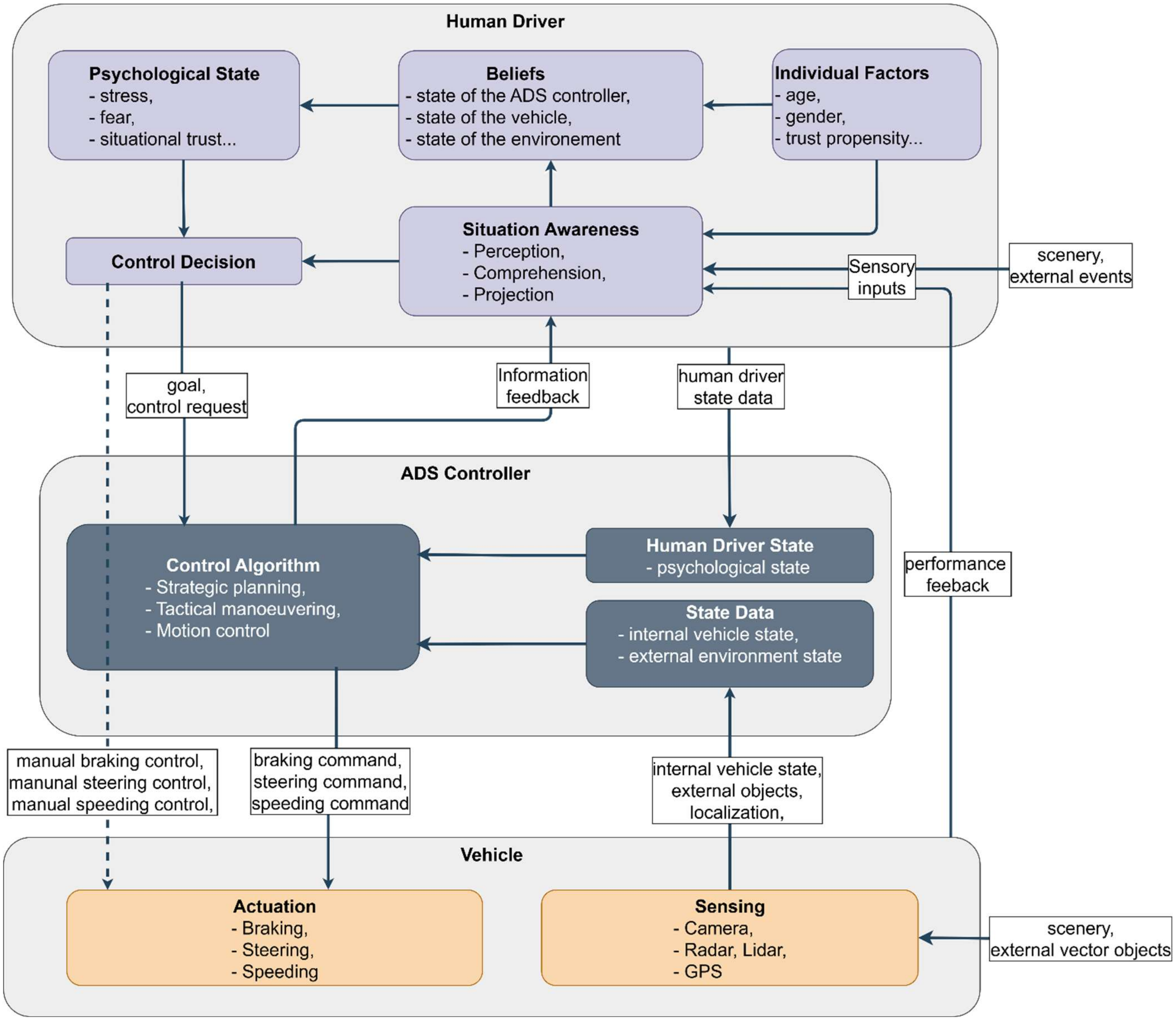

**Figure 5. Proposed psychological safety control structure.**

**Assign responsibilities (allocate psychological safety goals).** The next step of the STPA analysis is to assign responsibilities to elements of the system. The following lists examples of responsibilities assigned to the human driver and the ADS controller in the context of our illustrative scenario.

**Human driver responsibilities:**

- R1 - Understand the state of the DDT performance (assigned from SG1, SG2),
- R2 - Decide and request DDT takeover if needed (assigned from SG3).

**ADS controller responsibilities:**

- R3 - Respond to ADS Human driver requests (assigned from SG1, SG2),
- R4 - Command DDT control (assigned from SG3)
- R5 - Inform the human driver (assigned from SG3),
- R5 - Avoid unexpected behaviour (assigned from SG2),
- R6 - Monitor driver psychological state (assigned from SG2).

Note that responsibilities within the frame of STPA could be undertaken as a "Function" within the frame of standard systems engineering. In other words, the assignments of responsibilities as described in STAMP are equivalent to the derivation of safety goals into "Function Safety Requirements (FSR)" in ISO 26262.

Step3: Identify Unsafe Control Actions (UCAs)

The goal of the third step is to use the psychological safety control structure built previously to perform an analysis to identify psychologically Unsafe Control Actions (UCA) that could lead to psychological hazards in a given context. STAMP defines an UCA as "a control action that, in a particular context and worst-case environment, will lead to a hazard" (N. G. Leveson, 2012). The focus of our analysis being psychological safety, our UCAs are to be understood as control actions (including aspects of the control algorithm, interactions, control commands, and feedback) that will negatively impact the human driver's psychological state.

To help in the identification of UCAs, STPA provides the following 4 generic categories of UCAs that lead to hazards:

(1) not providing the control action,

(2) providing the control action (for instance unexpectedly, in excess, or too quickly),

(3) providing a potentially safe control action but too early, too late, or in the wrong order,

(4) providing a control action that lasts too long or is stopped too soon.

Note that by definition, a UCA occurs in a particular context and worst-case environment. Therefore, considering our example context, and without being exhaustive, we can formulate the following example UCAs utilizing the suggested categories:

- UCA1 - While experiencing a brake malfunction (unbalanced braking torque) during a lane change manoeuvre, ADS controller provides control commands (braking, steering) that cause vehicle to alternatively swerve left to right (H1).
- UCA2 - While mitigating a brake malfunction (unbalanced braking torque) during a lane change manoeuvre, ADS controller does not provide the human driver with enough clear information about the underlying issue (H5).
- UCA3 - While experiencing an unexpected behaviour (vehicle swerving left to right while changing lane), the ADS controller does not respond to the human driver's request to take over, leading to increasing the human driver's stress level (H2, H3).

Naturally, other driving scenarios can be assessed for relevant UCAs that can potentially lead to the psychological hazards considered for this analysis. These could include slightly different psychological variables (other passengers onboard, their age), different environmental conditions (climate, visibility etc.), or involve other critical driving manoeuvres (e.g. overtaking or emergency braking etc.). As a variation to our scenario, one could consider poor weather conditions (rain for instance) or the driver's the responsibility for protecting other passengers that would exacerbate or alter the observed malfunctioning behaviour and the psychological impact. For more completeness, scenarios in other driving contexts for UCAs leading to the same psychological hazards can be considered based on existing automated driving scenario taxonomy (*ISO 21448:2022 - Road Vehicles — Safety of the Intended Functionality*, n.d.) or derived from automated driving incident databases (*Standing General Order on Crash Reporting | NHTSA*, n.d.). These could involve different dynamic and scenery elements in country roads or inner-city driving environments.

Step 4: Identify Psychological loss scenarios

The last step identifies psychological loss scenarios which describe the causal factors within our previously defined system boundary that can lead to psychological hazards. STPA recommends considering two types of loss scenarios: 1) scenarios leading to unsafe control action occurrence and 2) scenarios leading to control actions not being executed or being improperly executed. STPA also provides four categories for causal factors that can explain why a controller might provide (or not provide) a control action that is unsafe. These categories include:

(5) failures involving the controller

(6) inadequate control algorithm

(7) unsafe control input

(8) inadequate process model

Considering two of the UCAs previously identified in our example use case illustration, Table 2 list four examples of psychological loss scenarios that can explain how the specified UCAs would occur.

Table 2. Examples of psychological loss scenarios leading to UCAs related to the illustrative example.

| **UCA** | **Responsibility** | **Scenario** |
|---|---|---|
| **UCA2** - While mitigating a brake malfunction (unbalanced braking torque) during a lane change manoeuvre, ADS controller does not provide the human driver with enough clear information about the underlying issue (H5). | Inform the human driver (SG3). | **UCA2.SC1 (inadequate control algorithm)** - During a lane change resulting in an alternative left to right vehicle motion, the ADS detects the malfunction but was not able to determine the problem's cause due to algorithmic limitations in tactical manoeuvre layer feature, causing the ADS controller's message to be generic. |
| | | **UCA2.SC2 (inadequate process model)** – During a lane change resulting in an alternative left to right vehicle motion, the ADS controller is unable to determine the cause of the problem due to incomplete internal vehicle sensor data from the braking subsystem, causing the ADS controller's message to be generic. |
| **UCA3** - While experiencing an unexpected behaviour (vehicle swerving left to | Respond to human driver | **UCA3.SC1 (inadequate control input)** – The ADS controller does not receive or receives the human driver's request too late due to faulty input from driver or inadequate priority arbitration in the control |
| right while changing lane), the ADS controller does not respond to the human driver's request to take over (H2, H3), increasing the human driver's stress level. | requests (SG2, SG3). | algorithm, leading to the request being ignored or delayed. |
| | | **UCA3.SC2 (inadequate control algorithm)** - The ADS controller received the human driver's request. But based on the operational situation (e.g. rain, speed, surrounding vehicles, traffic filtering etc), the ADS controller asserted that take over transition wouldn't leave enough time for the human driver to handle the situation appropriately. Thus, the ADS controller decided to handle the situation before giving control to the human driver. |

# 4 Discussion

Three main contributions identified in this paper include the definition of psychological safety in the context of AVs, its risk model, and analysis method based on STPA.

## *4.1 Psychological safety interpretation*

A new interpretation of psychological safety adapted for AVs was proposed, representing a new opportunity for autonomous systems safety research. By defining psychological safety in the AVs context, we can examine new impacts of interactions between AVs and passengers. In our example case, the vehicle performed in a physically safe manner (avoiding collision), but we find that its erratic behaviour limits driver belief that the system is performing safely and that their safety is ensured. This allows us to uncover a set of Unsafe Control Actions (UCAs) that can be avoided through psychologically safe system design, notably that the driver's unfamiliarity with the system's generic warning means that they are unable to recognize and mitigate malfunction of the braking system. While not investigated as the most threatening psychological losses in our example scenario, we also see how the psychological loss scenarios of fear and stress could relate to environment conditions, baseline psychological state, and system familiarity. Rainy weather and poor visibility could lead to automation distrust or the passengers could enter the vehicle in a state of high stress and susceptibility to stress response.

### 4.2 *Psychological safety risk model*

### 4.3 ***The risk model conceptualized the psychological harm impact mechanism in the context of human-AV interaction showing how the combination of system behaviour, environment factors, and the human-related individual factors such as psychological state or stake affect their appreciation of the system's performance. Furthermore, the risk model introduces a safety construct of psychological safety goals for mitigating psychological hazards and losses from which requirements specific to psychological safety can be derived. Such goals are allocated accordingly to system entities (ADS controller for instance) thanks to the responsibility assignment as described in the methodology. Finally, the risk model introduced the notion of PsySIL, adapted from ASILs and integrating a psychological severity scale as a way of appreciating the integrity required to avoid psychological losses. These concepts allow us to methodologically integrate psychological safety notions into the AV system design concept. Compared to existing psychological safety risk models described in the literature, our approach is proactive and systemic. Typically, all the previously described psychological models are applied to individuals and focus only on the severity aspect of psychological risk. Moreover, they are reactive in the sense that they are used to assess psychological harms that have already occurred, without rationale of how these harms occurred or how they can be prevented or mitigated.***

***In our approach the whole system is evaluated for contributing factors.***

***Psychological safety analysis method***

The first step of the proposed method introduced a new category of hazards and requirements for system design, showing the relevance and usefulness of the theoretical concepts defined earlier in the psychological safety risk model. The second step proposed a psychological control structure that enabled us to 1) capture the psychological dimension in the interaction between systems entities considering the driver's psychological state, situation awareness, and beliefs and 2) assign responsibilities to systems elements to prevent or mitigate the identified psychological hazard and losses. The responsibility assignment can be viewed as the formal allocation of psychological safety requirements to system entities, which is not currently done in current systems design practices. The third and fourth steps of the methodology identified the potential causes of the psychological hazards through UCAs, and psychological loss scenarios leading to these UCAs respectively. From these scenarios, it is possible to further refine psychological safety requirements for the system. For *UCA2.SC1* for instance, such further refinement could state the need for "the ADS controller to ensure explainability of messages sent to the human driver." Similarly, for *UCA3.SC2*, one could define the following additional high-level requirement*:* "The ADS controller must respect human agency and oversight." However, these new requirements can conflict with other requirements such as physical safety or cybersecurity. One way of addressing such challenges in systems analysis is to perform trade-off analyses which is, however, beyond the scope of this paper.

In this paper, we chose to use one simple scenario for the sake of being pedagogical. However, the approach is fully applicable to the entire range of possible driving scenarios that may pose psychological risk. In fact, identifying scenarios that causes psychological risk is one of the ultimate objectives of the proposed methodology, referring to the last step (identify psychological loss scenarios). Indeed, STPA itself, which is the technique our proposal is based on, is classified as a scenario-oriented safety analysis technique. Therefore, the use of a single illustrative example should not be seen as restrictive to the application of our proposal. As argued earlier, several taxonomies of automated driving scenarios and databases exist. They can be considered in order to widen the scope of the analysis. Also, slight variations in a given scenario can a be considered.

### *4.4 Limitations and future work*

The paper aimed to address a general problem posed by the emergence of psychological safety as a concern in humans and autonomous systems interaction. However, as specified throughout the paper, the scope of the definitions, proposals and analysis was focused on AVs. Therefore, more work is required to extend the proposals to autonomous systems in general, which opens avenues for future research.

Moreover, through the use of a single example scenario, it could be argued the scope of the analysis, which was limited to a SAE Level 4 autonomy in a highway driving context, was too narrow. However, the analysis can be extended to other scenarios with different psychological variables, environmental conditions, or critical driving manoeuvres. These scenarios can be based on existing automated driving scenario taxonomies (e.g., ISO 21448:2022) or derived from AVs incident databases that are publicly available. Although more examples, for instance involving inner city, country roads driving scenarios etc., could have been developed, this is a pedagogical choice motivated by the need for a simple example to ease the understanding of a proposed new idea without necessarily seeking a full validation (which could be done in future work).

Moreover, we acknowledge that even in the context of AVs, the proposal does not aim to replace the existing safety approaches (ISO 26262, SotIF); instead, it complements them by addressing potential psychological harm. It should also be acknowledged that psychological safety is to be managed in synergy with physical safety because the two concepts are inherently intertwined. Although not done in our current application, our framework offers the flexibility to allow consideration for both psychological and physical safety in the analysis. As described in Section 3, the analysis starts by defining the purpose of the analysis, where stakeholders' stakes are declined into losses which can derive from both psychological safety and physical safety stakes. However, prioritization could be required if certain stakes need to be prioritized or if there were to be conflicting requirements subsequent to the analysis. For such issues, trade-off analyses can be conducted on a case-by-case basis throughout the analysis to determine priority. Finally, psychological safety is an inherently broad concept that can encompass many of the previously investigated aspects of AV ridership such as trust and perceived control. However, future work should continue formalizing the psychological safety components to enable psychological safety to be dissected into easier-to-address sub-concepts. Additionally, our risk analysis falls far short of being able to estimate how individual experience can heighten the psychological risk of an event. For instance, it has been shown extensively that certain subpopulations tend to have higher levels of fear and concern towards AVs (Cugurullo & Acheampong, 2024; Thomas et al., 2020). To this end, our analysis is aimed at high level risk estimation, and we encourage any AV developer when using this analysis method to collaborate directly with psychologists to estimate the risk for certain subpopulations who might be system stakeholders.

## 5 Conclusion

This paper expressed the need to not only ensure physical safety of autonomous vehicles but also psychological safety. The paper defined the notion of psychological safety and proposed AV-PsySafe, a method of analysis to integrate psychological safety into the development process of AVs. In this context, the paper proposed to address psychological safety through an adaptation of STAMP and STPA. First, a risk model is defined for this need. The risk model defined and introduced the notion of psychological safety analysis in the autonomous systems interaction. In this risk model, PsySIL was proposed as a criterion for estimating AV psychological risk. Then an analysis method is proposed and applied step by step on the illustrative example to showcase its relevance and usefulness. We provide this proposal as a method that developers of autonomous vehicles can use to address psychological risk during the design, integration, and test phases of their development. Future work will attempt to address the identified limitation by generalizing the risk model to other autonomous systems, ensuring traceability of the analysis artefacts and ensuring the cohabitation of psychological safety with physical safety.

**Acknowledgement**

The authors would like to thank Dr Josimar Mendes from the University of Oxford for his review and feedback during the writing of this paper.

**Disclosure of interest**

The authors report there are no competing interests to declare.